\RequirePackage[british]{babel}
\pdfoutput=1
\documentclass[reqno,a4paper,12pt]{amsart}
\usepackage[a4paper,hmargin=2cm,tmargin=3cm,bmargin=3cm]{geometry}
\usepackage[utf8x]{inputenc}
\usepackage[T1]{fontenc}
\usepackage{mathptmx}
\usepackage{amsmath,amssymb,amstext,amsthm,amscd,mathrsfs,eucal,bm,slashed}
\usepackage{graphicx,color}
\usepackage{array}
\usepackage{cite}
\usepackage{hyperref}
\hypersetup{%
  pdftitle   = {Conformal Killing superalgebras},
  pdfkeywords = {rigid supersymmetry, conformal superalgebra, twistor spinors},
  pdfauthor  = {Paul de Medeiros},
  pdfcreator = {\LaTeX\ with package \flqq hyperref\frqq}
}
\PrerenderUnicode{éÉ}
\newcommand{\half}{\tfrac12}
\newcommand{\cB}{\mathcal{B}}
\newcommand{\cF}{\mathcal{F}}

\newcommand{\cL}{\mathcal{L}}

\newcommand{\cP}{\mathcal{P}}
\newcommand{\cR}{\mathcal{R}}
\newcommand{\cS}{\mathcal{S}}

\newcommand{\ff}{\mathfrak{f}}
\newcommand{\fg}{\mathfrak{g}}

\newcommand{\fB}{\mathfrak{B}}

\newcommand{\fF}{\mathfrak{F}}
\newcommand{\fG}{\mathfrak{G}}

\newcommand{\fR}{\mathfrak{R}}
\newcommand{\fS}{\mathfrak{S}}

\newcommand{\fX}{\mathfrak{X}}
\newcommand{\fgl}{\mathfrak{gl}}

\newcommand{\fk}{\mathfrak{k}}

\newcommand{\fp}{\mathfrak{p}}
\newcommand{\fr}{\mathfrak{r}}
\newcommand{\fs}{\mathfrak{s}}

\newcommand{\fso}{\mathfrak{so}}
\newcommand{\fosp}{\mathfrak{osp}}

\newcommand{\fsl}{\mathfrak{sl}}
\newcommand{\fsp}{\mathfrak{sp}}
\newcommand{\fsu}{\mathfrak{su}}
\newcommand{\fu}{\mathfrak{u}}

\newcommand{\Cl}{\mathrm{C}\ell}
\newcommand{\CCl}{\mathbb{C}\ell}
\newcommand{\SO}{\mathrm{SO}}
\newcommand{\Pin}{\mathrm{Pin}}
\newcommand{\Spin}{\mathrm{Spin}}

\newcommand{\PP}{\mathbb{P}}

\newcommand{\RR}{\mathbb{R}}
\newcommand{\CC}{\mathbb{C}}
\newcommand{\HH}{\mathbb{H}}
\newcommand{\KK}{\mathbb{K}}
\newcommand{\ZZ}{\mathbb{Z}}

\newcommand{\eT}{\mathscr{T}}

\DeclareMathOperator{\End}{End}

\DeclareMathOperator{\Mat}{Mat}

\DeclareMathOperator{\tr}{tr}

\newcommand{\rf}[1]{[\![#1]\!]}

\theoremstyle{plain}

\theoremstyle{definition}

\newcommand{\MUNCH}[1]{\relax}

\allowdisplaybreaks
\begin{document}
\title[Conformal symmetry superalgebras]{Conformal symmetry superalgebras}
\author[de Medeiros]{Paul de Medeiros}
\author[Hollands]{Stefan Hollands}
\address{School of Mathematics, Cardiff University, Senghennydd Road, Cardiff CF24 4AG, Wales, UK}
\email{demedeirospf@cardiff.ac.uk}
\address{Universit\"{a}t Leipzig, Institut f\"{u}r Theoretische Physik, Br\"{u}derstrasse 16, D-04103 Leipzig, FRG}
\email{HollandsS@cardiff.ac.uk}
\date{\today}
\vspace*{-.7cm}
\begin{abstract}
We show how the rigid conformal supersymmetries associated with a certain class of pseudo-Riemannian spin manifolds define a Lie superalgebra. The even part of this superalgebra contains conformal isometries and constant R-symmetries. The odd part is generated by twistor spinors valued in a particular R-symmetry representation. We prove that any manifold which admits a conformal symmetry superalgebra of this type must generically have dimension less than seven. Moreover, in dimensions three, four, five and six, we provide the generic data from which the conformal symmetry superalgebra is prescribed. For conformally flat metrics in these dimensions, and compact R-symmetry, we identify each of the associated conformal symmetry superalgebras with one of the conformal superalgebras classified by Nahm. We also describe several examples for Lorentzian metrics that are not conformally flat. 
\end{abstract}
\maketitle
\vspace*{-.7cm}
{\small{\tableofcontents}}

\clearpage

%%%%%%%%%%%
\section{Introduction}
\label{sec:introduction}

Supersymmetry multiplets may be either rigid or local, according to whether or not the dependence of the supersymmetry parameter on the background geometry is constrained. Local supermultiplets are coupled to supergravity with a dynamical background metric. By contrast, in the case of rigid supermultiplets, the metric is non-dynamical, and the supersymmetry parameter obeys a differential equation involving the metric and possibly other non-dynamical background fields. 

An important source of examples for constructing theories on curved backgrounds with rigid supersymmetry actually arises from supersymmetric backgrounds in supergravity. The constraint on the supersymmetry parameter here comes from setting to zero the supersymmetry variation of the fermionic fields in the supergravity multiplet. Beside their intrinsic interest, it is often possible to construct field theory multiplets with rigid supersymmetry on such backgrounds, as has been described in some detail in the recent literature \cite{Festuccia:2011ws,Jia:2011hw,Samtleben:2012gy,Klare:2012gn,Dumitrescu:2012he,Cassani:2012ri,Liu:2012bi,deMedeiros:2012sb,Dumitrescu:2012at,Kehagias:2012fh,Closset:2012ru,Martelli:2012sz,Samtleben:2012ua,Kuzenko:2012vd,Hristov:2013spa}.

A well-established feature of supersymmetric supergravity backgrounds is that they inherit a rigid Lie superalgebra structure, known as the {\emph{symmetry superalgebra}} of the background \cite{FigueroaO'Farrill:2004mx,FigueroaO'Farrill:2007ar,FOF:2007F4E8,FigueroaO'Farrill:2008ka,FigueroaO'Farrill:2008if}. The even part of this superalgebra contains the Killing vectors which generate isometries of the background. The odd part is generated by the rigid supersymmetries supported by the background. The image of the odd-odd bracket for the symmetry superalgebra spans a Lie subalgebra of Killing vectors for the background. This Lie subalgebra, together with the rigid supersymmetries, generate another rigid Lie superalgebra structure, called the {\emph{Killing superalgebra}}, which forms an ideal of the symmetry superalgebra. The utility of this construction is that it often allows one to infer important geometrical properties of the background directly from the rigid supersymmetry it supports. For example, in ten and eleven dimensions, it has recently been proven \cite{FigueroaO'Farrill:2012fp} that any supersymmetric supergravity background possessing more than half the maximal amount of supersymmetry is necessarily (locally) homogeneous.

The purpose of this note is to describe a similar construction for a class of backgrounds which support rigid conformal supersymmetry generated by {\emph{twistor spinors}}. This idea is not new and some previous attempts to define a Lie superalgebra structure for manifolds admitting twistor spinors can be found in \cite{Hab:1990,Klinker:2005,Rajaniemi:2006} (see also \cite{Duval:1993hs} for the construction of Schr\"{o}dinger superalgebras which do not involve twistor spinors). What distinguishes our construction is the inclusion of non-trivial R-symmetry which turns out to be crucial in order to solve the odd-odd-odd component of the Jacobi identity for the superalgebra. The defining condition for twistor spinors is conformally invariant and they furnish backgrounds on which rigid superconformal field theories may be defined, e.g. following the conformal coupling procedure described in \cite{deMedeiros:2012sb}.
\footnote{Our forthcoming paper \cite{dMHol2:2013}, which develops this construction for field theories with extended rigid supersymmetry, was in fact the original motivation for the present study.}
 Furthermore, at least in Euclidean and Lorentzian signatures, conformal equivalence classes of geometries admitting twistor spinors have been classified \cite{Baum:2002,BL:2003,Baum:2012,Leitner:2005,Baum:2008}. It is important to stress that the concept of twistor spinors we shall adhere to is entirely geometrical and unencumbered by the additional data (e.g. background fluxes and auxiliary fields) typically associated with a supergravity background. That is, we shall deal with what one might call \lq geometric' twistor spinors, for which their defining condition is with respect to the Levi-Civit\`{a} connection. It is possible to generalise our construction using twistor spinors defined with respect to a more elaborate superconnection for conformal supergravity backgrounds though we shall not explore that generalisation here.

Our aim is simply to ascribe a real Lie superalgebra to a conformal equivalence class of pseudo-Riemannian metrics on a spin manifold which admits a twistor spinor. This Lie superalgebra will be referred to as the {\emph{conformal symmetry superalgebra}} of the background. The even part of this superalgebra contains not only conformal Killing vectors for the background but also constant R-symmetries. The odd part is generated by twistor spinors valued in a particular representation of the R-symmetry. The image of the odd-odd bracket for the conformal symmetry superalgebra spans a Lie subalgebra of conformal Killing vectors and constant R-symmetries. This Lie subalgebra, together with the twistor spinors in the odd part, generate another rigid Lie superalgebra structure, that we will call the {\emph{conformal Killing superalgebra}}, which forms an ideal of the conformal symmetry superalgebra. We show that generic backgrounds admitting a conformal symmetry superalgebra must have dimension less than seven. Moreover, in dimensions three, four, five and six, we show that any manifold admitting twistor spinors generically inherits a conformal symmetry superalgebra with R-symmetry in a Lie subalgebra of a particular real form of $\fso_N$, $\fgl_N$, $\fsp_1$ and $\fsp_N$ respectively over $\CC$ (for some positive integer $N$). For the class of conformally flat metrics in these dimensions, if the R-symmetry Lie algebra is compact, we identify the conformal symmetry superalgebra with one of the conformal superalgebras classified by Nahm in \cite{Nahm:1977tg}. We shall exclude dimension two, where any pseudo-Riemannian metric is locally conformally flat, since then the associated conformal symmetry superalgebra is just as in flat space. We will also describe several examples of conformal Killing superalgebras for non-trivial conformal classes of Lorentzian metrics on manifolds that admit a twistor spinor. 

%%%%%%%%%%%
\section{Preliminaries}
\label{sec:preliminaries}

Let $M$ be a differentiable manifold (with real dimension $m$) equipped with pseudo-Riemannian metric $g$. The Levi-Civit\`{a} connection of $g$ will be denoted by $\nabla$. It will be assumed that $M$ has vanishing second Stiefel-Whitney class so the bundle $\SO(M)$ of oriented pseudo-orthonormal frames lifts to $\Spin (M)$ by the assignment of a spin structure.  

%%%%%%%%%%%
\subsection{Tensors, vectors and conformal Killing vectors}
\label{sec:TensorsVectorsCKV}

Let $\eT ( M )$ denote the space of tensor fields on $M$ (i.e. sections of the bundle of tensors over $M$). The Lie derivative $\cL_X$ along a vector field $X$ on $M$ defines an endomorphism of $\eT ( M )$. Such endomorphisms are induced by infinitesimal diffeomorphisms on $M$. For any $T , T^\prime \in \eT ( M )$, the Lie derivative obeys $\cL_X ( T + T^\prime ) = \cL_X T + \cL_X T^\prime$ and $\cL_X ( T \otimes T^\prime ) = \cL_X T  \otimes T^\prime + T \otimes \cL_X T^\prime$, for any vector field $X$ on $M$.    
 
Let $\fX ( M )$ denote the space of vector fields on $M$. The skewsymmetric bilinear map 
\begin{align}\label{eq:LieBracket}
[-,-] \; :\; \fX ( M ) \times \fX ( M ) \; &\rightarrow \; \fX ( M ) \nonumber \\
(X,Y) \; &\mapsto \; \cL_X Y = \nabla_X Y - \nabla_Y X~,
\end{align}
obeys the Jacobi identity and thus furnishes $\fX ( M )$ with the structure of a Lie algebra. The Lie derivative obeys identically
\begin{equation}\label{eq:LieDerHom}
[ \cL_X , \cL_Y ] = \cL_{[X,Y]}~,
\end{equation}
for all $X,Y \in \fX ( M )$ and so defines on $\eT ( M )$ a representation of the Lie algebra of vector fields. 

The subspace of conformal Killing vector fields in $\fX ( M )$ is given by
\begin{equation}\label{eq:CKV}
\fX^c ( M ) = \{ X \in \fX ( M ) \; |\; \cL_X g = -2 \sigma_X  g\}~,
\end{equation}
for some function $\sigma_X$ on $M$. Relative to a basis $\{ e_\mu \}$ on $\fX ( M )$, $\sigma_X = -\tfrac{1}{m} \nabla_\mu X^\mu$, for all $X \in \fX ( M )$. If $X,Y \in \fX^c ( M )$ then $[X,Y] \in \fX^c ( M )$. Whence, the restriction of the Lie bracket to $\fX^c ( M )$ defines a Lie subalgebra of conformal Killing vector fields on $M$. 

Any $X \in \fX^c ( M )$ with $\sigma_X =0$ is a Killing vector field and restricting the Lie bracket on $\fX^c ( M )$ to the subspace of Killing vector fields on $M$ defines a Lie subalgebra.  

%%%%%%%%%%%
\subsection{Clifford modules, spinors and twistor spinors}
\label{sec:Clifford algebrasSpinorsTS}

Let $\Cl (TM)$ denote the Clifford bundle over $M$ with defining relation 
\begin{equation}\label{eq:CliffordAlgebra}
{\bm X} {\bm Y} + {\bm X} {\bm Y} = 2 g(X,Y) {\bf 1}~,
\end{equation}
for all $X,Y \in \fX ( M )$. To each multivector field $\phi$ on $M$ there is an associated section ${\bm \phi}$ of $\Cl (TM)$. This reflects the fact that, at each point $x \in M$, the exterior algebra of $T_x M$ is isomorphic, as a vector space, to the Clifford algebra $\Cl (T_x M)$ (the metric $g$ and its inverse provide a duality between multivector fields and differential forms on $M$). With this isomorphism understood, we let $\{ {\bm e}_{\mu_1 ... \mu_k} | k=0,1,...,m  \}$ denote a basis for sections of $\Cl (TM)$ such that
\begin{equation}\label{eq:ClBasis}
{\bm e}_{\mu_1 ... \mu_k} = {\bm e}_{[ \mu_1} ... {\bm e}_{\mu_k ]} \equiv \frac{1}{k!} \sum_{\sigma \in S_k} (-1)^{|\sigma|} {\bm e}_{\mu_{\sigma(1)}} ... {\bm e}_{\mu_{\sigma(k)}}~,
\end{equation}
for degree $k>0$ (i.e. unit weight skewsymmetrisation of $k$ distinct degree one basis elements) and the identity element ${\bm 1}$ for $k=0$. The element ${\bm e}_{1...m}$ of maximal degree $m$ is proportional to an idempotent element ${\bm \Omega}$ that is associated with the volume form $\Omega$ for the metric $g$ on $M$. For $m$ odd, ${\bm \Omega}$ is central. For $m$ even, ${\bm \Omega} {\bm X} = - {\bm X} {\bm \Omega}$ for all $X \in \fX ( M )$.

The Clifford algebra $\Cl (T_x M)$ is $\ZZ_2$-graded such that elements with even and odd degrees are assigned grades $0$ and $1$ respectively. The grade $0$ elements span an ungraded associative subalgebra $\Cl^0 (T_x M) < \Cl (T_x M)$. The degree two elements span a Lie subalgebra $\fso (T_xM) < \Cl^0 (T_x M)$, where $\Cl^0 (T_x M)$ is understood as a Lie algebra whose brackets are defined by commutators. 

At each point $x \in M$, the set of invertible elements in $\Cl (T_x M)$ forms a multiplicative group $\Cl^\times (T_x M)$. The tangent vectors ${\bm X} \in \Cl (T_x M)$ with $g(X,X) = \pm 1$ generate the subgroup $\Pin ( T_x M ) < \Cl^\times (T_x M)$. The group $\Spin ( T_x M ) = \Pin ( T_x M ) \cap \Cl^0 (T_x M)$, which also follows by exponentiating $\fso (T_xM) < \Cl^0 (T_x M)$.

The {\emph{pinor}} module is defined by the restriction to $\Pin ( T_x M )$ of an irreducible representation of $\Cl (T_x M)$. Every Clifford algebra is isomorphic, as an associative algebra with unit, to a matrix algebra and it is a simple matter to deduce their irreducible representations. The {\emph{spinor}} module is defined by the restriction to $\Spin ( T_x M )$ of an irreducible representation of $\Cl^0 (T_x M)$. Note that restricting to $\Spin ( T_x M )$ an irreducible representation of $\Cl (T_x M)$ need not define an irreducible spinor module. For $m$ even, $\Cl (T_x M)$ has a unique irreducible representation which descends to a reducible representation when restricted to $\Spin ( T_x M )$, yielding a pair of inequivalent irreducible ({\emph{chiral}}) spinor modules associated with the two eigenspaces of ${\bm \Omega}$ on which ${\bm \Omega} = \pm 1$. For $m$ odd, $\Cl (T_x M)$ has two inequivalent irreducible representations which are isomorphic to each other when restricted to $\Spin ( T_x M )$. The isomorphism here is provided by the central element ${\bm \Omega}$ and corresponds to Hodge duality in the exterior algebra. In either case, the spinor module defined at each point in $M$ defines a principle bundle $\Spin (M)$ and its associated vector bundle is called the {\emph{spinor bundle}} over $M$.

Let $\fS ( M )$ denote the space of spinor fields on $M$ (i.e. sections of the spinor bundle over $M$). For $m$ even, $\fS ( M ) = \fS_+ (M) \oplus \fS_- (M)$, where $\fS_\pm ( M )$ denote the subspaces of chiral spinor fields on which ${\bm \Omega} = \pm 1$. The action of $\nabla$ induced on $\fS ( M )$ is compatible with the Clifford action, i.e.
\begin{equation}\label{eq:LCCliffComp}
\nabla_X ( {\bm Y} \psi ) = ( \nabla_X {\bm Y} )  \psi + {\bm Y} \nabla_X \psi~,
\end{equation}  
for all $X,Y \in \fX ( M )$ and $\psi \in \fS ( M )$. Furthermore
\begin{equation}\label{eq:LCBracket}
[ \nabla_X , \nabla_Y ] \psi = \nabla_{[X,Y]} \psi + \half {\bm R}(X,Y) \psi~,
\end{equation}
for all $X,Y \in \fX ( M )$ and $\psi \in \fS ( M )$, where ${\bm R}(X,Y) = \half X^\mu Y^\nu R_{\mu\nu\rho\sigma} {\bm e}^{\rho\sigma}$ in terms of the Riemann tensor $R_{\mu\nu\rho\sigma}$ of $g$. 

Now consider a map $\fp : \fX (M) \rightarrow \End \fS ( M )$ for which \\ [.2cm]
$\bullet$ $\fp$ is independent of the choice of pseudo-orthonormal frame on $M$. \\ [.1cm]
$\bullet$ $\fp_X ( f \psi ) = X(f) \psi + f \fp_X \psi$ , for all $X \in \fX ( M )$, $f \in C^\infty (M)$ and $\psi \in \fS ( M )$. \\ [.15cm]
$\bullet$ $\fp_X ( {\bm Y} \psi ) = {\bm{[X,Y]}}  \psi + {\bm Y} \fp_X \psi$, for all $X,Y \in \fX ( M )$ and $\psi \in \fS ( M )$. \\ [.2cm]   
It is readily verified that any such map is necessarily of the form
\begin{equation}\label{eq:SpinEnd}
\fp_X = \cL_X + \alpha_X~,
\end{equation}
acting on $\fS (M)$, for all $X \in \fX ( M )$, where
\begin{equation}\label{eq:SpinLieDer}
\cL_X = \nabla_X + \tfrac{1}{8} {\bm{dX^{\flat}}}~,
\end{equation}
and $\alpha_X$ is an arbitrary central element in the Clifford algebra. For any $X \in \fX ( M )$, $X^\flat$ denotes its dual one-form with respect to $g$, i.e. $X^\flat ( Y ) = g(X,Y)$ for all $Y \in \fX ( M )$. For any $X \in \fX^c (M)$, \eqref{eq:SpinLieDer} defines a {\emph{spinorial Lie derivative}} \cite{Lic:1963,Kosmann:1972,BG:1992,Habermann:1996}. 

It is useful to note that
\begin{align}\label{eq:LieDerBracketSpinors}
[ \cL_X , \nabla_Y ] \psi &= \nabla_{[X,Y]} \psi + \tfrac{1}{4} {\bm{d \sigma_X \!\wedge Y^\flat}} \psi \nonumber \\
[ \cL_X , {\bm \nabla} ] \psi & = \sigma_X {\bm \nabla} \psi - \left( \tfrac{m-1}{2} \right)  {\bm{\nabla \sigma_X}} \psi~,
\end{align}
for all $X \in \fX^c (M)$, $Y \in \fX ( M )$ and $\psi \in \fS ( M )$.

For any $X,Y \in \fX^c ( M )$, one finds that
\begin{equation}\label{eq:LieDerSpinHom}
[ \fp_X , \fp_Y ] = \fp_{[X,Y]}~,
\end{equation}
provided
\begin{equation}\label{eq:LieDerSpinHomAlpha}
\nabla_X \alpha_Y - \nabla_Y \alpha_X = \alpha_{[X,Y]}~.
\end{equation}
Whence, any $\alpha$ obeying \eqref{eq:LieDerSpinHomAlpha} defines on $\fS ( M )$ a representation of the Lie algebra of conformal Killing vector fields. A solution of \eqref{eq:LieDerSpinHomAlpha} is obtained by taking $\alpha_X$ proportional to $\sigma_X {\bf 1}$, for all $X \in \fX^c ( M )$. In particular, the representation defined by
\begin{equation}\label{eq:KSLieDer}
{\hat \cL}_X = \cL_X + \half  \sigma_X {\bf 1}~,
\end{equation}
for all $X \in \fX^c (M)$ is known as the {\emph{Kosmann-Schwarzbach Lie derivative}}.   

The subspace of conformal Killing (or twistor) spinor fields in $\fS ( M )$ is given by
\begin{equation}\label{eq:CKV}
\fS^c ( M ) = \{ \psi \in \fS ( M ) \; |\; \nabla_X \psi = \tfrac{1}{m} {\bm X} {\bm \nabla} \psi \; , \; \forall \, X \in \fX (M) \}~.
\end{equation}
Any $\psi \in \fS^c ( M )$ with ${\bm \nabla} \psi = \mu \psi$, for some constant $\mu$, is said to be {\emph{Killing}} if $\mu \neq 0$ or {\emph{parallel}} if $\mu =0$. The constant $\mu$ will be referred to as the {\emph{Killing constant}} of a Killing spinor $\psi$.

We define the {\emph{Penrose operator}} 
\begin{equation}\label{eq:CKV}
\cP_X = \nabla_X -  \tfrac{1}{m} {\bm X} {\bm \nabla}~,
\end{equation}
which acts on $\fS ( M )$ along any $X \in \fX (M)$. Whence, ${\mbox{ker}}\, \cP$ is precisely $\fS^c ( M )$. 

It is useful to note that
\begin{equation}\label{eq:LieDerBracketPenroseSpinors}
[ {\hat \cL}_X , \cP_Y ] \psi = \cP_{[X,Y]} \psi~,
\end{equation}
for all $X \in \fX^c (M)$, $Y \in \fX ( M )$ and $\psi \in \fS ( M )$. Whence, $\psi \in \fS^c ( M )$ implies ${\hat \cL}_X \psi \in \fS^c ( M )$ for all $X \in \fX^c (M)$.

It is also worth noting that the defining equation for twistor spinors implies 
\begin{equation}\label{eq:2DerTwistorSpinors}
\nabla_X {\bm \nabla} \psi  = \tfrac{m}{2} {\bm K}(X) \psi \;\; , \quad\quad {\bm \nabla}^2 \psi = - \tfrac{m}{4(m-1)} R\, \psi~,
\end{equation}
for all $X \in \fX (M)$ and $\psi \in \fS^c (M)$, where ${\bm K}(X) = X^\mu K_{\mu\nu} {\bm e}^\nu$ in terms of the Schouten tensor $K_{\mu\nu} = \tfrac{1}{m-2} \left( - R_{\mu\nu} + \tfrac{1}{2(m-1)} \, g_{\mu\nu} \, R \right)$, Ricci tensor $R_{\mu\nu}$ and scalar curvature $R$ of $g$. Combining \eqref{eq:2DerTwistorSpinors} with \eqref{eq:LCBracket} implies the integrability condition
\begin{equation}\label{eq:TwistorSpinorIntegrability}
{\bm W}(X,Y) \, \psi = 0~,
\end{equation}
for all $X,Y \in \fX (M)$ and $\psi \in \fS^c (M)$, where ${\bm W}(X,Y) = \half X^\mu Y^\nu W_{\mu\nu\rho\sigma} {\bm e}^{\rho\sigma}$ in terms of the Weyl tensor $W_{\mu\nu\rho\sigma} = R_{\mu\nu\rho\sigma} + g_{\mu\rho} K_{\nu\sigma} - g_{\nu\rho} K_{\mu\sigma} - g_{\mu\sigma} K_{\nu\rho} + g_{\nu\sigma} K_{\mu\rho}$.

%%%%%%%%%%%
\section{Conformal structure}
\label{sec:Weyl}

A Weyl transformation maps $g \mapsto \omega^2 g$, for some nowhere-vanishing function $\omega \in C^\infty (M)$. Any such Weyl transformation is compatible with the spin structure on $M$ provided ${\bm X} \mapsto \omega {\bm X}$ and $\psi \mapsto \omega^{\half} \psi$, for all $X \in \fX (M)$ and $\psi \in \fS (M)$. If a field $\Phi \mapsto \omega^{w_\Phi} \Phi$ under a Weyl transformation, for some $w_\Phi \in \RR$, it is said to have definite weight $w_\Phi$.

It is useful to note that the Penrose and Dirac operators do not transform with definite weight under Weyl transformations, but rather 
\begin{align}\label{eq:WeylPenroseDirac}
\cP_X &\mapsto \omega^{\half} \cP_X \omega^{-\half}  \nonumber \\ 
{\bm \nabla} &\mapsto \omega^{-\half (m+1)} {\bm \nabla}  \omega^{\half (m-1)}~,
\end{align}
for all $X \in \fX (M)$. The transformation of the Penrose operator in \eqref{eq:WeylPenroseDirac} implies that the space of twistor spinors $\fS^c (M)$ is Weyl-invariant. This means that the definition of $\fS^c (M)$ with respect to a given metric $g$ on $M$ with fixed spin structure extends trivially to the conformal equivalence class $[ g ]$ of metrics on $M$ that are related to $g$ by a Weyl transformation. It is also worth noting that the Christoffel symbols $\Gamma_{\mu\nu}^\rho = \half g^{\rho\sigma} ( \partial_\mu g_{\nu\sigma} + \partial_\nu g_{\mu\sigma} - \partial_\sigma g_{\mu\nu} )$ transform as 
\begin{equation}\label{eq:WeylChristoffel}
\Gamma_{\mu\nu}^\rho \mapsto \Gamma_{\mu\nu}^\rho + \omega^{-1} ( \delta^\rho_\mu \partial_\nu \omega + \delta^\rho_\nu \partial_\mu \omega - g_{\mu\nu} g^{\rho\sigma} \partial_\sigma \omega )~,
\end{equation}
under a Weyl rescaling of $g$. 

Using \eqref{eq:WeylPenroseDirac} and \eqref{eq:WeylChristoffel}, it follows that the spinorial Lie derivative in \eqref{eq:SpinLieDer} transforms such that
\begin{equation}\label{eq:WeylSpinLieDer}
\cL_X \psi \mapsto \omega^{\half} \cL_X \psi + \half \omega^{-\half} ( \partial_X \omega ) \psi~,
\end{equation}
for all $X \in \fX (M)$ and $\psi \in \fS^c (M)$. Consequently, the Kosmann-Schwarzbach Lie derivative in \eqref{eq:KSLieDer} transforms such that
\begin{equation}\label{eq:WeylKSLieDer}
{\hat \cL}_X \psi \mapsto \omega^{\half} {\hat \cL}_X \psi~,
\end{equation} 
for all $X \in \fX (M)$ and $\psi \in \fS^c (M)$. Whence $w_{{\hat \cL}_X \psi} = w_\psi = \half$ for any $X \in \fX (M)$. For $X \in \fX^c (M)$, this is as expected since ${\hat \cL}_X \psi \in \fS^c (M)$ from \eqref{eq:LieDerBracketPenroseSpinors}.

%%%%%%%%%%%
\section{Spinorial bilinear forms}
\label{sec:SpinBilinearForms}

Let $\langle -,- \rangle$ denote a non-degenerate bilinear form on $\fS (M)$ which obeys 
\begin{equation}\label{eq:SpinorInnerProd2}
X \langle \psi , \chi \rangle = \langle \nabla_X \psi , \chi \rangle + \langle \psi , \nabla_X \chi \rangle~,
\end{equation}
for all $X \in \fX (M)$ and $\psi , \chi \in \fS ( M )$. At any $x\in M$, \eqref{eq:SpinorInnerProd2} is equivalent to $\langle -,- \rangle$ being $\Spin ( T_x M )$-invariant. For $m$ even, this implies $\langle {\bm \Omega} \psi , \chi \rangle = (-1)^p \, \langle \psi , {\bm \Omega} \chi \rangle$, where $p = \lfloor \tfrac{m}{2} \rfloor$. Whence, 
\begin{align}\label{eq:ChiralSpinorProd}
\langle \psi_\pm , \chi_\mp \rangle &= 0 \quad {\mbox{if $p$ is even}} \nonumber \\ 
\langle \psi_\pm , \chi_\pm \rangle &= 0 \quad {\mbox{if $p$ is odd}}~,
\end{align}
for all $\psi_\pm , \chi_\pm \in \fS_\pm ( M )$. The projection operators ${\sf P}_\pm = \half ( {\bm 1} \pm {\bm \Omega} )$ define $\psi_\pm = {\sf P}_\pm \psi \in \fS_\pm ( M )$, for any $\psi \in \fS (M)$. 

To any pair $\psi , \chi \in \fS ( M )$, let us assign a vector field $\xi_{\psi , \chi}$ defined such that
\begin{equation}\label{eq:Bilinear}
g(X, \xi_{\psi , \chi} ) = \langle \psi , {\bm X} \chi \rangle~,
\end{equation}
for all $X \in \fX (M)$. From the results in section~\ref{sec:Weyl}, it follows that $\xi_{\psi , \chi}$ is Weyl-invariant. Furthermore, it is worth noting that 
\begin{equation}\label{eq:WeylNablaBilinear}
\langle \psi , {\bm \nabla} \chi \rangle \mapsto  \langle \psi , {\bm \nabla} \chi \rangle + \frac{m}{2} \, \omega^{-1} \partial_{\xi_{\psi , \chi}} \omega~,
\end{equation}
under a Weyl transformation, for all $\psi , \chi \in \fS ( M )$.

It is useful to note that
\begin{equation}\label{eq:LieDerBilinear}
\cL_X \xi_{\psi , \chi} = \xi_{{\hat \cL}_X \psi , \chi} + \xi_{\psi , {\hat \cL}_X \chi} ~,
\end{equation}
for all $X \in \fX^c (M)$ and any $\psi , \chi \in \fS ( M )$, in terms of the Kosmann-Schwarzbach Lie derivative on the right hand side. Furthermore, it follows that $\xi_{\psi , \chi} \in \fX^c (M)$ if $\psi , \chi \in \fS^c (M)$.

Now let the dual ${\overline \psi}$ of any $\psi \in \fS (M)$ with respect to $\langle -,- \rangle$ be defined such that ${\overline \psi} \chi = \langle \psi , \chi \rangle$, for all $\chi \in \fS (M)$. If $\psi , \chi \in \fS ( M )$ then $\psi {\overline \chi} \in \End \fS ( M )$. At each point $x \in M$, we may express any such endomorphism of the spinor module in terms of the action of $\Cl( T_x M)$, relative to the basis defined in \eqref{eq:ClBasis}. This expression is called a {\emph{Fierz identity}} and follows using the identities 
\begin{equation}\label{eq:FierzTrace}
\tr ( {\bm e}_{\mu_1 ... \mu_k} {\bm e}^{\nu_1 ... \nu_l} ) = \begin{cases} (-1)^{\lfloor \tfrac{k}{2} \rfloor}  k! \, \delta_{[ \mu_1}^{\nu_1} ... \delta_{\mu_k ]}^{\nu_k} \tr ( {\bf 1} ) & \quad {\text{if $k=l$}} \\
0 & \quad {\text{if $k\neq l$}}~,
\end{cases}
\end{equation}
involving the canonical trace form $\tr$ for the matrix representation. 

For $m$ odd, this yields
\begin{equation}\label{eq:moddFierz}
\psi {\overline \chi} = \frac{1}{2^p} \sum_{k=0}^p \frac{(-1)^{\lfloor \tfrac{k}{2} \rfloor}}{k!} ( {\overline \chi} {\bm e}^{\mu_1 ... \mu_k} \psi ) {\bm e}_{\mu_1 ... \mu_k}~,
\end{equation}
for all $\psi , \chi \in \fS ( M )$.

For $m = 0$ mod $4$, this yields
\begin{align}\label{eq:mzeromodfourFierz}
\psi_\pm {\overline \chi}_\pm &= \frac{1}{2^{p-1}} \! \left( \sum_{k=0}^{\tfrac{p}{2} -1} \frac{ (-1)^k}{(2k)!}  ( {\overline \chi}_\pm {\bm e}^{\mu_1 ... \mu_{2k}} \psi_\pm ) {\bm e}_{\mu_1 ... \mu_{2k}} +  \frac{(-1)^{\tfrac{p}{2}}}{2\, p!}  ( {\overline \chi}_\pm {\bm e}^{\mu_1 ... \mu_{p}} \psi_\pm ) {\bm e}_{\mu_1 ... \mu_{p}} \right) \! {\sf P}_\pm \nonumber \\ 
\psi_\pm {\overline \chi}_\mp &= \frac{1}{2^{p-1}} \sum_{k=0}^{\tfrac{p}{2} -1}  \frac{(-1)^k}{(2k+1)!}  ( {\overline \chi}_\mp {\bm e}^{\mu_1 ... \mu_{2k+1}} \psi_\pm ) {\bm e}_{\mu_1 ... \mu_{2k+1}} {\sf P}_\mp~,
\end{align}
for all $\psi_\pm , \chi_\pm \in \fS_\pm ( M )$.

For $m = 2$ mod $4$, this yields
\begin{align}\label{eq:mtwomodfourFierz}
\psi_\pm {\overline \chi}_\pm &= \frac{1}{2^{p-1}} \! \left( \sum_{k=0}^{\tfrac{p-3}{2}} \frac{ (-1)^k}{(2k+1)!}  ( {\overline \chi}_\pm {\bm e}^{\mu_1 ... \mu_{2k+1}} \psi_\pm ) {\bm e}_{\mu_1 ... \mu_{2k+1}} + \frac{(-1)^{\tfrac{p-1}{2}} }{2\, p!}  ( {\overline \chi}_\pm {\bm e}^{\mu_1 ... \mu_{p}} \psi_\pm ) {\bm e}_{\mu_1 ... \mu_{p}} \right) \! {\sf P}_\mp \nonumber \\ 
\psi_\pm {\overline \chi}_\mp &= \frac{1}{2^{p-1}} \sum_{k=0}^{\tfrac{p-1}{2}} \frac{(-1)^k}{(2k)!}  ( {\overline \chi}_\mp {\bm e}^{\mu_1 ... \mu_{2k}} \psi_\pm ) {\bm e}_{\mu_1 ... \mu_{2k}} {\sf P}_\pm~,
\end{align}
for all $\psi_\pm , \chi_\pm \in \fS_\pm ( M )$.

Spin-invariant bilinear forms on $\fS ( M )$ have been classified \cite{Harvey:1990,AlekCort1995math,Alekseevsky:2003vw} and their type depends on both the dimension $m$ and the signature of $g$. To facilitate our understanding of these types, we need first to review a few more details concerning the classification of Clifford algebras and their spinor representations. We begin with the simpler case of complex Clifford algebras before discussing their real forms. See \cite{Harvey:1990} for a more detailed account of these results. 

We shall employ the following notation in our description of the relevant modules. Let $\KK (N) = \Mat_N ( \KK )$ denote the associative algebra of $N{\times}N$ matrices with entries in a field $\KK = \RR, \CC, \HH$, and let $2\KK (N) = \KK (N) \oplus \KK (N)$. The matrix algebra $\KK (N)$ defines a Lie algebra $\fgl_N (\KK)$, with Lie bracket defined by the matrix commutator. We will refer to the representation of a Lie algebra as being of {\emph{type}} $\KK$ if it can be realised in terms of a matrix algebra over ground field $\KK$ acting on a $\KK$-vector space.    

%%%%%%%%%%%
\subsection{Complex case}
\label{sec:ComplexCase}

Let us define $\CCl (TM) = \Cl (T_\CC M)$, as the Clifford bundle associated with the complexified tangent bundle $T_\CC M = TM \otimes_\RR (M{\times}\CC)$. At each point $x \in M$, $\CCl (T_x M)$ is isomorphic to the complexification of $\Cl (T_x M)$. This complex Clifford algebra will be denoted by $\CCl (m)$. The complex-bilinear extension of $g$ required to define $\CCl (TM)$ renders its signature immaterial. 

A fundamental isomorphism for complex Clifford algebras is $\CCl (m+2) \cong \CCl (m) \otimes \CCl (2)$. Since $\CCl (0) \cong \CC$, $\CCl (1) \cong 2 \CC$ and $\CCl (2) \cong \CC (2)$, it follows that
\begin{equation}\label{eq:ComplexClifford}
\CCl (m) \cong \begin{cases}
\CC (2^p) & {\text{if $m$ is even}} \\ 
2 \CC (2^p) & {\text{if $m$ is odd}}~.
\end{cases}
\end{equation}
The pinor module is therefore
\begin{equation}\label{eq:ComplexCliffordPinor}
\PP \cong \begin{cases}
\CC^{2^p} & {\text{if $m$ is even}} \\ 
2\CC^{2^p} & {\text{if $m$ is odd}}~.
\end{cases}
\end{equation}

Another important isomorphism is $\CCl^0 (m+1) \cong \CCl (m)$, as ungraded associative algebras. Whence,
\begin{equation}\label{eq:ComplexEvenClifford}
\CCl^0 (m) \cong \begin{cases}
2 \CC (2^{p-1}) & {\text{if $m$ is even}} \\ 
\CC (2^p) & {\text{if $m$ is odd}}~.
\end{cases}
\end{equation}
The spinor module  is therefore
\begin{equation}\label{eq:ComplexCliffordSpinor}
{\mathbb{S}} \cong \begin{cases}
2\CC^{2^{p-1}} & {\text{if $m$ is even}} \\ 
\CC^{2^p} & {\text{if $m$ is odd}}~.
\end{cases}
\end{equation}
For $m$ even, ${\mathbb{S}}$ is reducible and it reduces to the direct sum of two irreducible chiral spinor modules ${\mathbb{S}}_\pm \cong \CC^{2^{p-1}}$. The irreducible modules ${\mathbb{S}}_\pm$ correspond to the two eigenspaces of the idempotent element ${\bm \Omega} = i^p \, {\bm e}_{1...m} \in \CCl(m)$, with respective eigenvalues $\pm 1$.  

It can be shown that every spin-invariant bilinear form ${\sf C}$ on ${\mathbb{S}}$ obeys 
\begin{equation}\label{eq:ComplexCliffordBilinear}
{\sf C}( \psi , \chi ) = \sigma_{\sf C} \, {\sf C}( \chi , \psi ) \; , \quad\quad  {\sf C}( {\bm X} \psi , \chi ) = \tau_{\sf C} \, {\sf C}( \psi , {\bm X} \chi )~,
\end{equation}
for all $\psi , \chi \in {\mathbb{S}}$ and $X \in \CC^m$, in terms of some fixed pair of signs $\sigma_{\sf C}$ and $\tau_{\sf C}$. From the second condition in \eqref{eq:ComplexCliffordBilinear}, it follows that ${\sf C}( {\bm \Omega} \psi , \chi ) = \tau_{\sf C}^m (-1)^p \, {\sf C}( \psi , {\bm \Omega} \chi )$. For $m$ odd, this implies $\tau_{\sf C} = (-1)^p$. For $m$ even, in addition to ${\sf C}$, one can define a new spin-invariant bilinear form ${\sf C}^\prime (-,-) = {\sf C} (-,{\bm \Omega}-)$ with $\sigma_{{\sf C}^\prime} = (-1)^p \sigma_{\sf C}$ and $\tau_{{\sf C}^\prime} = - \tau_{\sf C}$. Whence, a spin-invariant bilinear form with $\tau = -1$ can always be chosen if $m$ is even. Any such bilinear form has the virtue of being $\CCl(m)$-invariant and we shall always select this one when $m$ is even. For any $m$, this fixes $\sigma = (-1)^{\tfrac{p(p+1)}{2}}$. Henceforth, the properties in \eqref{eq:ComplexCliffordBilinear} with the aforementioned sign choices for $\sigma$ and $\tau$ will be ascribed to $\langle -,- \rangle$. 

 %%%%%%%%%%%
\subsection{Real case}
\label{sec:RealCase}

Let us assume that $g$ has $s$ positive and $t$ negative eigenvalues at each point $x \in M$. Thus $m=s+t$ and let $n=t-s$ denote the signature of $g$. The Clifford algebra $\Cl (T_x M)$ will be denoted by $\Cl (s,t)$. Note that $\Cl (s,t) \otimes_\RR \CC \cong \CCl (m)$.

Three fundamental isomorphisms for real Clifford algebras are 
\begin{align}\label{eq:RealCliffordIso}
\Cl(s+2,0) &\cong \Cl(0,s) \otimes \Cl(2,0) \nonumber \\ 
\Cl(s+1,t+1) &\cong \Cl(s,t) \otimes \Cl(1,1) \nonumber \\
\Cl(0,t+2) &\cong \Cl(t,0) \otimes \Cl(0,2)~.
\end{align}
Since $\Cl (0) \cong \RR$, $\Cl (1,0) \cong 2 \RR$, $\Cl (0,1) \cong \CC$,  $\Cl (2,0) \cong \RR(2) \cong \Cl (1,1)$ and  $\Cl (0,2) \cong \HH$, it follows that
\begin{equation}\label{eq:RealClifford}
\Cl (s,t) \cong \begin{cases}
\RR (2^p) & {\text{if $n =0,6$ mod $8$}} \\
2 \RR (2^p) & {\text{if $n =7$ mod $8$}} \\
\CC (2^p) & {\text{if $n =1,5$ mod $8$}} \\
\HH (2^{p-1}) & {\text{if $n =2,4$ mod $8$}} \\ 
2 \HH (2^{p-1}) & {\text{if $n =3$ mod $8$}}~.
\end{cases}
\end{equation}

The matrix algebra isomorphisms $\KK (N) \otimes_\RR \RR (N^\prime) \cong \KK (N N^\prime) $, $\CC \otimes_\RR \CC \cong 2 \CC$, $\CC \otimes_\RR \HH \cong  \CC (2)$ and $\HH \otimes_\RR \HH \cong \RR (4)$ are useful in deriving this classification of real Clifford algebras. The pinor module is therefore
\begin{equation}\label{eq:RealCliffordPinor}
{\mathrm{P}} \cong \begin{cases}
\RR^{2^p} & {\text{if $n =0,6$ mod $8$}} \\
2 \RR^{2^p} & {\text{if $n =7$ mod $8$}} \\
\CC^{2^p} & {\text{if $n =1,5$ mod $8$}} \\
\HH^{2^{p-1}} & {\text{if $n =2,4$ mod $8$}} \\ 
2 \HH^{2^{p-1}} & {\text{if $n =3$ mod $8$}}~.
\end{cases}
\end{equation}

Two further important ungraded isomorphisms are $\Cl^0 (s+1,t) \cong \Cl^0 (s,t+1)  \cong \Cl (s,t)$. Whence,
\begin{equation}\label{eq:RealEvenClifford}
\Cl^0 (s,t) \cong \begin{cases}
\RR (2^p) & {\text{if $n =1,7$ mod $8$}} \\
2 \RR (2^{p-1}) & {\text{if $n =0$ mod $8$}} \\
\CC (2^{p-1}) & {\text{if $n =2,6$ mod $8$}} \\
\HH (2^{p-1}) & {\text{if $n =3,5$ mod $8$}} \\ 
2 \HH (2^{p-2}) & {\text{if $n =4$ mod $8$}}~.
\end{cases}
\end{equation}
From \eqref{eq:RealEvenClifford}, we infer another useful isomorphism $\Cl^0 (s,t) \cong \Cl^0 (t,s)$ which implies that spinor representations of real Clifford algebras do not depend on the sign of $n$. The spinor module  is
\begin{equation}\label{eq:RealCliffordSpinor}
{\mathrm{S}} \cong \begin{cases}
\RR^{2^p} & {\text{if $n =1,7$ mod $8$}} \\
2 \RR^{2^{p-1}} & {\text{if $n =0$ mod $8$}} \\
\CC^{2^{p-1}} & {\text{if $n =2,6$ mod $8$}} \\
\HH^{2^{p-1}} & {\text{if $n =3,5$ mod $8$}} \\ 
2 \HH^{2^{p-2}} & {\text{if $n =4$ mod $8$}}~.
\end{cases}
\end{equation}

If $m$ is even then so is $n$. If $n=0$ mod $4$ then ${\mathrm{S}}$ is reducible and it reduces to the direct sum of two irreducible chiral spinor modules ${\mathrm{S}}_\pm$. If $n =0$ mod $8$, then ${\mathrm{S}}_\pm \cong \RR^{2^{p-1}}$ is irreducible. If $n =4$ mod $8$, then ${\mathrm{S}}_\pm \cong \HH^{2^{p-2}}$ is irreducible. If $n=2$ mod $4$, then ${\mathrm{S}} \cong \CC^{2^{p-1}}$ is irreducible. In this case, ${\bm e}_{1...m}$ defines a complex structure on ${\mathrm{P}} _\CC$ and the submodule on which ${\bm e}_{1...m} =i$ is isomorphic to ${\mathrm{S}}$.

If $m$ is odd then so is $n$. If $n =1,7$ mod $8$, then ${\mathrm{S}} \cong \RR^{2^{p}}$ is irreducible. If $n =3,5$ mod $8$, then ${\mathrm{S}} \cong \HH^{2^{p-1}}$ is irreducible. 

The results of \cite{AlekCort1995math,Alekseevsky:2003vw} show that every spin-invariant bilinear form ${\sf c}$ on ${\mathrm{S}}$ obeys 
\begin{equation}\label{eq:RealCliffordBilinear}
{\sf c}( \psi , \chi ) = \sigma_{\sf c} \, {\sf c}( \chi , \psi ) \; , \quad\quad  {\sf c}( {\bm X} \psi , \chi ) = \tau_{\sf c} \, {\sf c}( \psi , {\bm X} \chi )~,
\end{equation}
for all $\psi , \chi \in {\mathrm{S}}$ and $X \in \RR^{s,t}$, in terms of some fixed pair of signs $\sigma_{\sf c}$ and $\tau_{\sf c}$. The classification in \cite{AlekCort1995math,Alekseevsky:2003vw} proceeds by establishing a bijection between isomorphism classes of spin-invariant bilinear forms on ${\mathrm{S}}$ and elements in the Schur algebra ${\mathscr{C}} ( {\mathrm{S}} )$ of $\fso(s,t)$-invariant endomorphisms of ${\mathrm{S}}$. These Schur algebras are isomorphic to the matrix algebras displayed in Table~\ref{tab:tableRealSchur} and depend only on $n$ mod $8$.     
\begin{table}
\begin{center}
\begin{tabular}{|c||c|c|c|c|c|c|c|c|}
  \hline
  $n$ & $0$ & $1$ & $2$ & $3$ & $4$ & $5$ & $6$ & $7$ \\ 
  \hline
  &&&&&&&& \\ [-.4cm]
  ${\mathscr{C}} ( {\mathrm{S}} )$ & $2\RR$ & $\RR(2)$ & $\CC(2)$ & $\HH$ & $2\HH$ & $\HH$ & $\CC$ & $\RR$  \\ \hline
  &&&&&&&& \\ [-.4cm]
  ${\mbox{dim}}_\RR {\mathscr{C}} ( {\mathrm{S}} )$ & $2$ & $4$ & $8$ & $4$ & $8$ & $4$ & $2$ & $1$  \\   
  \hline 
\end{tabular} \vspace*{.2cm}
\caption{Schur algebras ${\mathscr{C}} ( {\mathrm{S}} )$ and their real dimensions.}
\label{tab:tableRealSchur}
\end{center}
\end{table}
In the complex case, the Schur algebras are isomorphic to either $2\CC$ for $m$ even or $\CC$ for $m$ odd, corresponding to the classes of invariant bilinear forms ${\sf C}$ described at the end of section~\ref{sec:ComplexCase}. For each complex spin-invariant bilinear form ${\sf C}$ on ${\mathbb{S}}$ there is a corresponding real spin-invariant bilinear form ${\sf c}$ on ${\mathrm{S}}$ with the same symmetry properties. Therefore, it will be convenient to assume the same sign choices made for $\langle -,- \rangle$ in both the real and complex cases. Note however that Table~\ref{tab:tableRealSchur} illustrates there are several alternative options for real spin-invariant bilinear forms on ${\mathrm{S}}$ with different symmetry properties that we shall not concern ourselves with.  

%%%%%%%%%%%
\section{Conformal symmetry superalgebras}
\label{sec:CKS}

Let $\cS = \cB \oplus \cF$ denote the $\ZZ_2$-graded real vector space on which we shall define a Lie superalgebra structure. The even part $\cB = \fX^c ( M ) \oplus \cR$, where $\cR$ is a real Lie algebra with constant parameters on $M$. The complexification of the odd part $\cF$ is isomorphic to either 
\begin{align}\label{eq:OddComplexification}
\fS^c (M) \otimes_\CC W &\quad {\mbox{if $m$ is odd}} \nonumber \\
\fS^c_+ (M) \otimes_\CC V \oplus \fS^c_- (M) \otimes_\CC V^* &\quad {\mbox{if $m=0$ mod $4$}} \nonumber \\
\fS^c_+ (M) \otimes_\CC W &\quad {\mbox{if $m=2$ mod $4$}}~,
\end{align}
where $V$ and $W$ are certain complex $\cR_\CC$-modules. $V^*$ denotes the dual module of $V$. $W$ admits a (skew)symmetric $\cR_\CC$-invariant nondegenerate bilinear form $b$, which provides an isomorphism $W^* \cong W$. We shall assume that $\cS$ is finite-dimensional and therefore take $m > 2$. 

%%%%%%%%%%%
\subsection{Brackets}
\label{sec:brackets}

The graded Lie bracket on $\cS$ is a bilinear map $[-,-] : \cS \times \cS \rightarrow \cS$, defined such that 
\begin{equation}\label{eq:SuperBracket}
[\cB,\cB] \subset \cB \;\; , \quad\quad [\cB,\cF] \subset \cF \;\; , \quad\quad [\cF,\cF] \subset \cB~.
\end{equation}
These component brackets are defined as follows. 

$\bullet$ The skewsymmetric $[\cB,\cB]$ bracket is given by $[ \fX^c (M) , \fX^c (M) ] \oplus [ \cR , \cR ]$, in terms of the Lie bracket of conformal Killing vector fields on $M$ and the Lie bracket on $\cR$. The mixed $[ \fX^c (M) , \cR ]$ contribution is absent since elements in $\cR$ are constant on $M$.

$\bullet$ The skewsymmetric $[\cB,\cF]$ bracket is defined such that 
\begin{equation}\label{eq:evenoddbracket}
[ X , \epsilon ] = {\hat \cL}_X \epsilon \;\; , \quad\quad [ \rho , \epsilon ] = \rho \cdot \epsilon~,
\end{equation}
for all $X \in \fX^c (M)$, $\rho \in \cR$ and $\epsilon \in \cF$. In the second bracket, $\cdot$ denotes the $\cR$-action of $\rho$. 

$\bullet$ The symmetric $[\cF,\cF]$ bracket is bilinear in its entries. Whence, one needs only to specify the bracket of any given element in $\cF$ with itself in order to extract the general $[\cF,\cF]$ bracket. That is, knowing $[ \epsilon , \epsilon ]$ for all $\epsilon \in \cF$ allows one to deduce $[ \epsilon , \epsilon^\prime ]$, for any $\epsilon , \epsilon^\prime \in \cF$, via the polarisation $\half \left( [ \epsilon + \epsilon^\prime , \epsilon + \epsilon^\prime ] - [ \epsilon , \epsilon ] - [ \epsilon^\prime , \epsilon^\prime ] \right)$. In this way, the $[\cF,\cF]$ bracket is defined generically by 
\begin{equation}\label{eq:oddoddbracket}
[ \epsilon , \epsilon ] = \xi_\epsilon + \rho_\epsilon~,
\end{equation}
in terms of certain elements $\xi_\epsilon \in \fX^c (M)$ and $\rho_\epsilon \in \cR$ which we will now define. 

Let $\Xi_\epsilon$ be defined such that
\begin{equation}\label{eq:XiComplex}
g( X , \Xi_\epsilon ) = \begin{cases} 
b_{ij} \, {\overline \epsilon}^i {\bm X} \epsilon^j & {\text{if $m$ is odd}} \\
{\overline \epsilon}_+^i {\bm X} \epsilon_{-\, i} & {\text{if $m=0$ mod $4$}} \\
b_{ij} \, {\overline \epsilon}_+^i {\bm X} \epsilon_+^j & {\text{if $m=2$ mod $4$}}~,
\end{cases}
\end{equation}
for all $X \in \fX_\CC (M)$, where $\epsilon \in \cF_\CC$, relative to a basis $\{ e_i \}$ for either $V$ or $W$. A dual basis $\{ e^i \}$ for either $V^*$ or $W^*$ is defined such that $e^i ( e_j) = \delta^i_j$, with $e_i = b_{ij} e^j$ for $W \cong W^*$. Using \eqref{eq:LieDerBilinear}, one finds that $\Xi_\epsilon \in \fX^c_\CC (M)$. Moreover, $\Xi_\epsilon$ is clearly $\cR_\CC$-invariant. The component $\xi_\epsilon$ in \eqref{eq:oddoddbracket} is defined as the real part of $\Xi_\epsilon$.

Let $\Pi_\epsilon \in \cR_\CC$ be defined such that
\begin{equation}\label{eq:PiComplex}
( \Pi_\epsilon \cdot \psi )^i = \begin{cases} 
\alpha \, ( {\overline \epsilon}^i {\bm \nabla} \epsilon_j - \kappa \, {\overline \epsilon}_j{\bm \nabla} \epsilon^i ) \psi^j & {\text{if $m$ is odd}} \\
\alpha \, ( {\overline \epsilon}_+^i {\bm \nabla} \epsilon_{-\, j} - \kappa\, {\overline \epsilon}_{-\, j} {\bm \nabla} \epsilon_+^i ) \psi^j + \beta \, ( {\overline \epsilon}_+^j {\bm \nabla} \epsilon_{-\, j} - \kappa\, {\overline \epsilon}_{-\, j} {\bm \nabla} \epsilon_+^j ) \psi^i & {\text{if $m=0$ mod $4$}} \\
\alpha \, ( {\overline \epsilon}_+^i {\bm \nabla} \epsilon_{+\, j} - \kappa \, {\overline \epsilon}_{+\, j} {\bm \nabla} \epsilon_+^i ) \psi^j & {\text{if $m=2$ mod $4$}}~,
\end{cases}
\end{equation}
where $\kappa$ is the sign for which ${\overline \psi} {\bm X} \chi = \kappa \, {\overline \chi} {\bm X} \psi$, for all $X \in \fX_\CC (M)$ and $\psi , \chi \in \cF_\CC$. The values of constants $\alpha$ and $\beta$ will be determined in a moment. Using \eqref{eq:2DerTwistorSpinors}, one finds that $\Pi_\epsilon$ is constant. Moreover, the $\cR_\CC$-action on $W$ defined by the first and third lines of \eqref{eq:PiComplex} preserves $b$. The component $\rho_\epsilon$ in \eqref{eq:oddoddbracket} is defined as the real part of $\Pi_\epsilon$.

Now let $\rho_{\epsilon,\epsilon^\prime} = \half ( \rho_{\epsilon + \epsilon^\prime} - \rho_\epsilon - \rho_{\epsilon^\prime} )$, for any $\epsilon , \epsilon^\prime \in \cF$. It is useful to note that the second identity in \eqref{eq:LieDerBracketSpinors} implies 
\begin{equation}\label{eq:rhoLie}
\rho_{{\hat \cL}_X \epsilon , \epsilon^\prime} + \rho_{\epsilon , {\hat \cL}_X \epsilon^\prime} =0~,
\end{equation}
for all $X \in \fX^c (M)$. Furthermore, 
\begin{equation}\label{eq:rhoLie2}
[ \rho , \rho_{\epsilon,\epsilon^\prime} ] =  \rho_{\rho \cdot \epsilon,\epsilon^\prime} + \rho_{\epsilon,\rho \cdot \epsilon^\prime}~,
\end{equation}
for all $\rho \in \cR$.

%%%%%%%%%%%
\subsection{Conformal invariance}
\label{sec:conformalinvariance}

Under a Weyl transformation, the results of section~\ref{sec:Weyl} imply that $\cB$ and $\cF$ must transform with definite weights $w_{\cB} = 0$ and $w_{\cF} = \half$. These weight assignments are compatible with the graded Lie brackets defined above. Compatibility is trivial for the $[\cB,\cB]$ bracket. For the $[\cB,\cF]$ bracket, it follows using \eqref{eq:WeylKSLieDer}. For the $[\cF,\cF]$ bracket, it follows by checking that $\xi_\epsilon$ and $\rho_\epsilon$, defined in \eqref{eq:XiComplex} and \eqref{eq:PiComplex}, are Weyl-invariant. For $\xi_\epsilon$, this is obvious. For $\rho_\epsilon$, it follows using \eqref{eq:WeylNablaBilinear}. Whence, the conformal symmetry superalgebra $\cS$ should be ascribed to a conformal equivalence class of metrics on $M$ with fixed spin structure.

%%%%%%%%%%%
\subsection{Jacobi identities}
\label{sec:JacobiIdentities}

If $\cS$ is a Lie superalgebra then the brackets defined above must obey the graded Jacobi identity. There are four distinct graded components, of type $[\cB\cB\cB]$, $[\cB\cB\cF]$, $[\cB\cF\cF]$ and $[\cF\cF\cF]$, each of which must vanish identically.

$\bullet$ The $[\cB\cB\cB]$ component vanishes trivially since both $\fX^c ( M )$ and $\cR$ are Lie algebras.

$\bullet$ The $[\cB\cB\cF]$ component vanishes using ${\hat \cL}_{[X,Y]} = [ {\hat \cL}_X , {\hat \cL}_Y ]$, for all $X,Y \in \fX^c (M)$, and that $V$ and $W$ are $\cR_\CC$-modules.

$\bullet$ The $[\cB\cF\cF]$ component contains two parts. The part in $\fX^c (M)$ vanishes using \eqref{eq:LieDerBilinear} together with the fact that $\xi_\epsilon$ is $\cR$-invariant. The part in $\cR$ vanishes using \eqref{eq:rhoLie} and \eqref{eq:rhoLie2}. 

$\bullet$ The $[\cF\cF\cF]$ component is totally symmetric and trilinear in $\cF$. Whence, via polarisation, it is equivalent to the condition
\begin{equation}\label{eq:oddoddoddjacobi}
{\hat \cL}_{\xi_\epsilon} \epsilon + \rho_\epsilon \cdot \epsilon = 0~,
\end{equation}
for all $\epsilon \in \cF$. Unlike the first three, this final component is non-trivial but can be solved case by case using the Fierz identities \eqref{eq:moddFierz}, \eqref{eq:mzeromodfourFierz} and \eqref{eq:mtwomodfourFierz}, the symmetry properties of $\langle -,- \rangle$ described in section~\ref{sec:ComplexCase} together with the definitions \eqref{eq:XiComplex} and \eqref{eq:PiComplex}. It is also necessary to make use of the identity 
\begin{equation}\label{eq:Gamma2Identity}
{\bm e}^{\nu\rho} {\bm e}_{\mu_1 ... \mu_k} {\bm e}_{\nu\rho} = ( m - (m-2k)^2 ) {\bm e}_{\mu_1 ... \mu_k}~,
\end{equation}
in this calculation.

The result is as follows. If $m\geq 7$, one finds that \eqref{eq:oddoddoddjacobi} has no generic solutions for any choice of $\alpha$ and $\beta$. In this context, \lq generic' just means that \eqref{eq:oddoddoddjacobi} is solved without assuming further special properties for twistor spinors on $M$. We shall encounter several non-trivial conformal classes of Lorentzian metrics which admit non-generic solutions with $m\geq 7$ in section~\ref{sec:examples}. The generic solutions of \eqref{eq:oddoddoddjacobi} for $2 < m < 7$ are summarised in Table~\ref{tab:tableCKS}. In each case, $\cR_\CC  < \fG$ for a particular type of complex reductive Lie algebra $\fG$. For $m=4$, if $N = {\mbox{dim}}_\CC V$, the component of $\cR_\CC$ in the centre $Z ( \fgl ( V ) ) \cong \CC$ acts with coefficient $\half \tfrac{4-N}{4N}$ in \eqref{eq:PiComplex}, whence $\cR_\CC < \fsl ( V )$ if $N=4$. For $m=3$, if ${\mbox{dim}}_\CC W =1$, then \eqref{eq:oddoddoddjacobi} is solved for any value of $\alpha$.
\begin{table}
\begin{center}
\begin{tabular}{|c|c|c|c|c|}
  \hline
  $m$ & $\alpha$ & $\beta$ & $\fG$ & comment \\ 
  \hline \hline
   &  &  &  &  \\ [-.4cm] 
  $6$ & $\tfrac{2}{3}$ & $*$ & $\fsp (W)$ &  \\ [.1cm] 
  $5$ & $\tfrac{3}{5}$ & $*$ & $\fsp (W)$ & ${\mbox{dim}}_\CC W =2$ \\ [.1cm]
  $4$ & $\tfrac{1}{2}$ & $-\tfrac{1}{8}$ & $\fgl (V)$ & ${\mbox{dim}}_\CC V \neq 4$ \\ [.1cm]
  $4$ & $\tfrac{1}{2}$ & $-\tfrac{1}{8}$ & $\fsl (V)$ & ${\mbox{dim}}_\CC V =4$ \\ [.1cm] 
  $3$ & $\tfrac{2}{3}$ & $*$ & $\fso (W)$ & ${\mbox{dim}}_\CC W \neq 1$  \\ [.1cm] 
  \hline 
\end{tabular} \vspace*{.2cm}
\caption{Generic data for complexified conformal symmetry superalgebras.}
\label{tab:tableCKS}
\end{center}
\end{table}

%%%%%%%%%%%
\subsection{Real forms}
\label{sec:RealForms}

Let us conclude this section by providing a more detailed account of the real structure of conformal symmetry superalgebras. Relative to the complexified construction described above, this involves taking a real form $\cR$ of $\cR_\CC$ and identifying a real structure on $\cF_\CC$. We shall consider only those real forms $\cS$ with $\cF$ in a real representation of $\cB$. At each point in $M$, $\cF_\CC$ defined in \eqref{eq:OddComplexification} involves the tensor product (over $\CC$) of a complex irreducible spinor module and an $\cR_\CC$-module. To define a real structure on $\cF_\CC$ requires the irreducible spinor module and the $\cR$-module in $\cF$ to be of the same type $\KK$. 

The real form $\cR$ is a Lie subalgebra of some real form $\fg$ of $\fG$. Where possible, we shall take $\fg$ to be the compact real form of $\fG$. Table~\ref{tab:SSRealForm} lists all the real forms $\fk$ of the relevant complex semisimple Lie algebras $\fk_\CC$, following the nomenclature in chapter 26 of \cite{FulHar}. In each case, the type of the defining representation of $\fk$ appears in the third column. The compact real forms in Table~\ref{tab:SSRealForm} (with $l=0$) will be written $\fsp(k)$, $\fsu(k)$ and $\fso(k)$.   
\begin{table}
\begin{center}
\begin{tabular}{|c|c|c|c|}
  \hline
  $\fk_\CC$ & $\fk$ & type & alias    \\ [.05cm]
  \hline\hline
  &&& \\ [-.4cm]
  $\fsp_k (\CC)$ & $\fsp_k (\RR)$ & $\RR$ &    \\ [.05cm]
  &&& \\ [-.4cm]
  $\fsp_{k+l} (\CC)$ & $\fu_{k,l} (\HH)$ & $\HH$ & $\fsp (k,l)$    \\ [.05cm]
  &&& \\ [-.4cm]
  $\fsl_k (\CC)$ & $\fsl_k (\RR)$ & $\RR$ &    \\ [.05cm]
  &&& \\ [-.4cm]
  $\fsl_{k+l} (\CC)$ & $\fsu_{k,l}$ & $\CC$ &  $\fsu (k,l)$  \\ [.05cm]
  &&& \\ [-.4cm]
  $\fsl_{2k} (\CC)$ & $\fsl_k (\HH)$ & $\HH$ & $\fsu^* (2k)$   \\ [.05cm]
  &&& \\ [-.4cm]
  $\fso_{k+l} (\CC)$ & $\fso_{k,l} (\RR)$ & $\RR$ &  $\fso (k,l)$  \\ [.05cm]
  &&& \\ [-.4cm]
  $\fso_{2k} (\CC)$ & $\fu_k^* (\HH)$ & $\HH$ & $\fso^* (2k)$   \\ [.05cm]
  \hline  
\end{tabular} \vspace*{.2cm}
\caption{Real forms of complex semisimple Lie algebras.}
\label{tab:SSRealForm}
\end{center}
\end{table}

From the classification of real spinor modules in \eqref{eq:RealCliffordSpinor} and the real forms in Table~\ref{tab:SSRealForm}, it is straightforward to match up irreducible spinor modules with $\fg$-modules of the same type. The generic data for the corresponding real conformal symmetry superalgebras in $2 < m < 7$  is summarised in Table~\ref{tab:RealCKV}. Signatures $(s,t)$ and $(t,s)$ give rise to isomorphic real forms so we have taken $s\geq t$ in Table~\ref{tab:RealCKV}. 
\begin{table}
\begin{center}
\begin{tabular}{|c|c|c|c|}
  \hline
  $m$ & $(s,t)$ & $\fg$ & type   \\ 
  \hline \hline
   &  &  &   \\ [-.4cm]
  $6$ & $(3,3)$ & $\fsp_N (\RR)$ & $\RR$    \\ [.1cm] 
  $6$ & $(5,1)$ & $\fsp (N)$ & $\HH$   \\ [.1cm] 
  $5$ & $(3,2)$ & $\fsp_1 (\RR)$ & $\RR$   \\ [.1cm]
  $5$ & $(4,1)$ & $\fsp (1)$ & $\HH$   \\ [.1cm]
  $5$ & $(5,0)$ & $\fsp (1)$ & $\HH$   \\ [.1cm]
  $4$ & $(2,2)$ & $\fgl_{N\neq 4} (\RR)$ & $\RR$   \\ [.1cm]
  $4$ & $(2,2)$ & $\fsl_4 (\RR)$ & $\RR$  \\ [.1cm] 
  $4$ & $(3,1)$ & $\fu (N\neq 4)$ & $\CC$   \\ [.1cm] 
  $4$ & $(3,1)$ & $\fsu (4)$ & $\CC$  \\ [.1cm] 
  $4$ & $(4,0)$ & $\fu^* ( 2N\neq 4)$ & $\HH$ \\ [.1cm] 
  $4$ & $(4,0)$ & $\fsu^*(4)$ & $\HH$   \\ [.1cm] 
  $3$ & $(2,1)$ & $\fso (N)$ & $\RR$   \\ [.1cm]
  $3$ & $(3,0)$ & $\fso^* (2N)$ & $\HH$  \\ [.1cm] 
  \hline 
\end{tabular} \vspace*{.2cm}
\caption{Generic data for real conformal symmetry superalgebras.}
\label{tab:RealCKV}
\end{center}
\end{table}
Note that it is only possible to take $\fg$ compact in Euclidean ($t=0$) and Lorentzian ($t=1$) signatures. However, in Euclidean signature, $\fg$ is generically noncompact if $m<5$. In the $(4,0)$ case, $\fg$ is always noncompact though, if $N=1$, only the centre $Z ( \fgl_1 (\HH) ) \cong \fso(1,1)$ is noncompact since $\fsl_1 (\HH) \cong \fsp(1)$. In the $(3,0)$ case, $\fg$ is noncompact unless $N=1$, in which case $\fu_1^* (\HH) \cong \fu(1)$.  

%%%%%%%%%%%
\section{Comparison with Nahm for conformally flat metrics}
\label{sec:Nahm}

Finite-dimensional classical Lie superalgebras over $\CC$ have been classified by Kac
\cite{Kac:1975} (see also \cite{Scheunert:1976uf,Scheunert:1976ug}). Now consider the subclass of all such Lie superalgebras $\fS$ which contain an $\fso_n ( \CC )$ factor in their even part $\fB$ and a spinor representation of $\fso_n ( \CC )$ in their odd part $\fF$. In each case, $\fB = \fso_n ( \CC ) \oplus \fR$, where $\fR$ is a reductive Lie algebra over $\CC$. For $n>4$, the complete list of all non-isomorphic Lie superalgebras of this type is given in Table~\ref{tab:tableKac}. Entries in the \lq $\fF$' column denote tensor product representations of $\fB$. The first factor corresponds to an irreducible spinor representation of $\fso_n ( \CC )$  and the second factor corresponds to the defining representation of $\fR$. In the \lq $\fF$' column entry for ${\mathfrak{sl}}_{4 | N\neq 4} (\CC)$, the centre of $\fgl_N (\CC)$ acts on $\fF$ with weight $4-N$. 

\begin{table}
\begin{center}
\begin{tabular}{|c|c|c|c|}
  \hline
  $n$ & $\fS$ & $\fR$ & $\fF$   \\ 
  \hline \hline
   &  &  &    \\ [-.4cm]
  $8$ & $\fosp_{8 | N} (\CC)$ & $\fsp_N (\CC)$ & ${\mathbb{S}}_+ \otimes \CC^{2N}$   \\ [.1cm] 
  $7$ & ${\mathfrak{f}}_4 (\CC)$ & $\fsp_1 (\CC)$ & ${\mathbb{S}} \otimes \CC^{2}$    \\ [.1cm]
  $6$ & ${\mathfrak{sl}}_{4 | N\neq 4} (\CC)$ & $\fgl_{N} (\CC)$ & ${\mathbb{S}}_+ \otimes \CC^{N}$  \\ [.1cm] 
   $6$ & ${\mathfrak{psl}}_{4|4} (\CC)$ & $\fsl_4 (\CC)$ & ${\mathbb{S}}_+ \otimes \CC^{4}$   \\ [.1cm] 
  $5$ & $\fosp_{N | 2} (\CC)$ & $\fso_N (\CC)$ & ${\mathbb{S}} \otimes \CC^{N}$  \\ [.1cm]
  \hline 
\end{tabular} \vspace*{.2cm}
\caption{Classical complex Lie superalgebras involving spinor representations.}
\label{tab:tableKac}
\end{center}
\end{table}

Real forms of the classical Lie superalgebras in \cite{Kac:1975} have been classified by Parker \cite{Parker:1980}. Up to isomorphism, the real forms of each complex classical Lie superalgebra are uniquely determined by the real forms of the complex reductive Lie algebra which constitutes it even part. It is therefore straightforward to deduce all the real forms ${\mathfrak{s}}$ of a given complex Lie superalgebra $\fS$ of the type defined in the paragraph above. Now consider all such real Lie superalgebras for which  
 
$\bullet$ The real form ${\mathfrak{r}}$ of $\fR$ is of compact type.

$\bullet$ The real form ${\mathfrak{f}}$ of $\fF$ is in a real representation of the real form ${\mathfrak{b}}$ of $\fB$.

For $4< n \leq 8$, the relevant real forms of  $\fso_n ( \CC )$ are isomorphic to $\fso (k,l)$, for some $k+l=n$. Real forms ${\mathfrak{b}}$ with ${\mathfrak{r}}$ compact may therefore be indexed by pairs $(k,l)$. For ${\mathfrak{f}}$ to be in a real representation of ${\mathfrak{b}}$ requires that the representation type of both $\fso (k,l)$ and ${\mathfrak{r}}$ factors must be the same. That is, they must both be of the same type $\KK$ in order to equip ${\mathfrak{f}}$ with a real structure.

For $n>4$, the complete list of all non-isomorphic real Lie superalgebras meeting the two conditions above is given in Table~\ref{tab:tableNahm}. It is precisely this classification that was obtained by Nahm in \cite{Nahm:1977tg}. The real forms associated with pairs $(k,l)$ and $(l,k)$ give rise to isomorphic real Lie superalgebras and so we present them with $k > l$. Entries in the \lq ${\mathfrak{f}}$' column denote particular real forms of tensor product (over $\CC$) representations of ${\mathfrak{b}}$ which are defined as follows. Given a pair of quaternionic representations $W$ and $W^\prime$, equipped with quaternionic structures $J$ and $J^\prime$, then $J {\otimes} J^\prime$ defines a real structure on $W \otimes W^\prime$ and $[ W \otimes W^\prime ]$ denotes the real representation induced on the fixed points of $J {\otimes} J^\prime$ on $W \otimes W^\prime$. On the other hand, given a complex representation $V$, $\rf{V}$ denotes the real representation obtained by restricting scalars from $\CC$ to $\RR$. 
\begin{table}
\begin{center}
\begin{tabular}{|c|c|c|c|c|c|c|}
  \hline
  $n$ & $\fs$ & $\fr$ & $\ff$ & $(k,l)$ & type & Nahm label  \\ 
  \hline \hline
   &  &  &  & & & \\ [-.4cm]
  $8$ & $\fosp(6,2|N)$ & $\fsp (N)$ & $[ {\mathrm{S}}_+ \otimes \HH^N ]$ & $(6,2)$ & $\HH$ & {\tt X} \\ [.1cm] 
  $7$ & $\ff(4)^{\prime\prime}$ & $\fsp (1)$ & $[ {\mathrm{S}} \otimes \HH ]$ & (5,2) & $\HH$ & {\tt IX}$_2$  \\ [.1cm]
  $7$ & $\ff(4)^\prime$ & $\fsp (1)$ & $[ {\mathrm{S}} \otimes \HH ]$ & (6,1) & $\HH$ &  {\tt IX}$_1$ \\ [.1cm]
  $6$ & $\fsu(2,2|N\neq 4)$ & $\fu (N)$ & $\rf{{\mathrm{S}}_+ \otimes \CC^N}$ & $(4,2)$ & $\CC$ &  {\tt VIII} \\ [.1cm] 
  $6$ & ${\mathfrak{psu}}(2,2|4)$ & $\fsu (4)$ & $\rf{{\mathrm{S}}_+ \otimes \CC^4}$ & $(4,2)$ & $\CC$ &  {\tt VIII}$_1$ \\ [.1cm] 
  $6$ & $\fsu(4|N \neq 4)$ & $\fu (N)$ & $\rf{{\mathrm{S}}_+ \otimes \CC^N}$ & $(6,0)$ & $\CC$ & {\tt VIII}$^\prime$ \\ [.1cm] 
  $6$ & ${\mathfrak{psu}}(4|4)$ & $\fsu (4)$ & $\rf{{\mathrm{S}}_+ \otimes \CC^4}$ & $(6,0)$ & $\CC$ & {\tt VIII}$^\prime_1$ \\ [.1cm] 
  $5$ & $\fosp_{N|2}(\RR)$ & $\fso (N)$ & ${\mathrm{S}} \otimes \RR^N$ & (3,2) & $\RR$ & {\tt VII} \\ [.1cm]
  $5$ & $\fosp (2|1,1)$ & $\fu (1)$ & $[ {\mathrm{S}} \otimes \HH ]$ & (4,1) & $\HH$ & {\tt VII}$_1$ \\ [.1cm]
  $5$ & $\fosp(2|2)$ & $\fu (1)$ & $[ {\mathrm{S}} \otimes \HH ]$ & (5,0) & $\HH$ & {\tt VII}$^\prime_1$ \\ [.1cm]
  \hline 
\end{tabular} \vspace*{.2cm}
\caption{Real Lie superalgebras with $n\, {>}\, 4$ in Nahm's classification.}
\label{tab:tableNahm}
\end{center}
\end{table}

Let us now review the interpretation of some of these real Lie superalgebras as conformal superalgebras. The Lie algebra of conformal isometries of $\RR^{s,t}$ is isomorphic to $\fso(s+1,t+1)$. Therefore, the three real Lie superalgebras {\tt VIII}$^\prime$, {\tt VIII}$^\prime_1$ and {\tt VII}$^\prime_1$ in Table~\ref{tab:tableNahm} cannot be associated with conformal superalgebras for $\RR^{s,t}$. Moreover, those which remain can only be identified with conformal superalgebras for $\RR^{s,t}$ in either Euclidean ($l=1$) or Lorentzian ($l=2$) signature. In Euclidean signature, only two real Lie superalgebras {\tt IX}$_1$, {\tt VII}$_1$ in Table~\ref{tab:tableNahm} can describe conformal superalgebras in dimensions $s+t =5,3$. In Lorentzian signature, the remaining five real Lie superalgebras {\tt X}, {\tt IX}$_2$, {\tt VIII} ({\tt VIII}$_1$) and {\tt VII} in Table~\ref{tab:tableNahm} describe the conformal superalgebras for Minkowski space in dimensions $s+t =6,5,4,3$. On any Riemannian or Lorentzian spin manifold $M$ with conformally-flat metric $g$ in dimension $m>2$, each real conformal symmetry superalgebra $\cS$ in Table~\ref{tab:RealCKV} with $\fg$ compact is isomorphic to one of the conformal superalgebras in Table~\ref{tab:tableNahm}. 

%%%%%%%%%%%
\section{Some non-trivial examples in Lorentzian signature}
\label{sec:examples}

The general structure of conformal isometries for manifolds admitting a twistor spinor is extremely complicated. Thus, rather than attempting to classify conformal symmetry superalgebras, we shall instead conclude with some non-trivial examples  for Lorentzian manifolds which admit a twistor spinor. Before doing so, let us first briefly dispatch the Euclidean case in the paragraph below.

Up to local conformal equivalence, the classification of Riemannian manifolds $M$ with $m\geq 3$ which admit a nowhere-vanishing twistor spinor $\epsilon$ is conceptually straightforward. The results of \cite{Lich:1988,BFGK:1991} imply that each conformal class of metrics on any such $M$ contains a unique representative that is Einstein with constant non-negative scalar curvature. If the constant scalar curvature is zero then $\epsilon$ is parallel. The classification of all complete simply connected irreducible non-flat Riemannian manifolds which admit a parallel spinor is due to Wang \cite{Wang:1989}. If the constant scalar curvature is positive then $\epsilon$ is a Killing spinor. The Killing constant of $\epsilon$ is either real or imaginary. The classification of all complete simply connected irreducible non-flat Riemannian manifolds which admit a Killing spinor with real/imaginary Killing constant is due to Baum \cite{Baum:1989}/B\"{a}r \cite{Bar:1993}. Whence, conformal Killing superalgebras for Riemannian manifolds admitting a twistor spinor correspond to Killing superalgebras generated by Killing spinors, but for the inclusion of a non-trivial R-symmetry. 

For any Lorentzian spin manifold $M$, there exists a spin-invariant non-degenerate pseudo-Hermitian inner product $(-,-)$ on $\fS (M)$, which obeys 
\begin{equation}\label{eq:HermitianInnerProd}
( \psi , \chi )^* = ( \chi , \psi )  \;\; , \quad\quad ( {\bm X} \psi , \chi ) = - ( \psi , {\bm X} \chi )
\end{equation}
for all $\psi , \chi \in \fS (M)$ and $X \in \fX (M)$, where $*$ denotes complex conjugation. 

In general, this inner product is distinct from the bilinear form $\langle -,- \rangle$ defined in section~\ref{sec:SpinBilinearForms}. However, in cases where the complex spinor bundle on $M$ admits a complex-antilinear automorphism ${\sf B}$ which squares to $\pm 1$ (i.e. a real or quaternionic structure), then one can identify $\langle -,- \rangle$ with $( - , {\sf B} -)$ provided ${\sf B}$ is compatible with the pseudo-Hermitian structure defined by $(-,-)$. This requires $\langle {\bm X} \psi , \chi \rangle = - \langle \psi , {\bm X} \chi \rangle$, for all $X \in \fX (M)$ and $\psi , \chi \in \fS (M)$. From section~\ref{sec:SpinBilinearForms}, we recall that one can always choose a bilinear form with this property on the complex spinor bundle except when $m = 1$ mod $4$. 

To any $\psi \in \fS (M)$, one can assign a vector field $\zeta_{\psi}$ defined such that
\begin{equation}\label{eq:DiracCurrent}
g(X, \zeta_{\psi} ) = -i ( \psi , {\bm X} \psi )~,
\end{equation}
for all $X \in \fX (M)$. The vector field $\zeta_\psi$ is known as the {\emph{Dirac current}} of $\psi$. It is readily verified that $\zeta_\psi \in \fX^c (M)$ if $\psi \in \fS^c (M)$. From \eqref{eq:HermitianInnerProd}, it follows that $\zeta_\psi$ is real and has strictly non-positive norm. Moreover, $\zeta_\psi$ is null only if ${\bm \zeta_\psi} \psi =0$ and $( \psi , \psi )=0$. It is worth noting that the Dirac current $\zeta_\psi$ is nowhere-vanishing only if the same is true of $\psi$.

Let us also assign the function 
\begin{equation}\label{eq:DiracConstant}
\varsigma_{\psi} = ( \psi , {\bm \nabla} \psi )~,
\end{equation}
to any $\psi \in \fS (M)$. If $\psi \in \fS^c (M)$ and $\varsigma_{\psi}$ is real then it is also constant on $M$, using \eqref{eq:HermitianInnerProd} and the first identity in \eqref{eq:2DerTwistorSpinors}. If $\zeta_\psi$ is a Killing vector then $\varsigma_{\psi}$ is real. 

Up to local conformal equivalence, the classification of simply connected complete Lorentzian manifolds $M$ with $m \geq 3$ which admit a nowhere-vanishing twistor spinor $\epsilon$ can be found in \cite{Leitner:2005,Baum:2008}. 

If $\epsilon$ is parallel then $M$ can always be decomposed into the direct product of irreducible non-flat Riemannian manifolds (each of which must admit a parallel spinor) and a certain Lorentzian manifold $L$. If $\zeta_\epsilon$ is timelike, then $L$ is locally isometric to Minkowski space. If $\zeta_\epsilon$ is null, then $L$ is locally isometric to a {\emph{Brinkmann-wave}}. 

If $\epsilon$ is not parallel then $M$ is locally isometric to either 

$\bullet$ A Lorentzian {\emph{Einstein-Sasaki}} manifold. \\ [.1cm]
$\bullet$ A {\emph{Fefferman space}}. \\ [.1cm]
$\bullet$ The direct product of a Lorentzian Einstein-Sasaki manifold and an irreducible non-flat Riemannian manifold which admits a Killing spinor (with imaginary Killing constant). 
 
Let us now consider each of these cases in turn.

%%%%%%%%%%%
\subsection{Brinkmann waves}
\label{sec:Brinkmannwaves}

By definition, a Brinkmann-wave $L$ is a Lorentzian manifold which admits a parallel null vector field $v$. This vector field defines a flag of subbundles $\RR v \subset v^\perp \subset TL$, where $v^\perp = \{ X \in \fX(L) \, |\, g(X,v) =0 \}$. The metric induced from $g$ on the bundle $E = v^\perp / \RR v$ is Riemannian and let ${\mathrm{Hol}}_\nabla (E)$ denote the holonomy group of the associated Levi-Civit\`{a} connection on $E$. If $L$ has dimension $l$ then, at each point in $L$, ${\mathrm{Hol}}_\nabla (E)$ is a Lie subgroup of $\Spin(l-2)$ which preserves a spinor on $E$. The holonomy group ${\mathrm{Hol}}_\nabla (L)$ of the Levi-Civit\`{a} connection for a Brinkmann-wave $L$ which admits a parallel spinor is isomorphic to ${\mathrm{Hol}}_\nabla (E) \ltimes \RR^{l-2}$. A Brinkmann-wave with ${\mathrm{Hol}}_\nabla (E)$ trivial is called a {\emph{pp-wave}}.

Any twistor spinor $\epsilon$ on a Brinkmann-wave is necessarily parallel and obeys ${\bm v} \epsilon =0$. For any pair $\epsilon , \epsilon^\prime \in \fS^c (L)$, the vector field $\xi_{\epsilon , \epsilon^\prime}$ (defined by \eqref{eq:Bilinear}) is parallel and orthogonal to $v$. Given another null vector field $u$ with $g(u,v) =1$, which always exists locally, it follows that $g( X , \xi_{\epsilon , \epsilon^\prime} ) = g( X,v) g( \xi_{\epsilon , \epsilon^\prime} , u )$ for any $X \in \fX (L)$, using $\half ( {\bm u} {\bm v} + {\bm v} {\bm u} ) =  {\bf 1} $ and that $\epsilon$ and $\epsilon^\prime$ are in the kernel of ${\bm v}$. Consequently, $\xi_{\epsilon , \epsilon^\prime} = g( \xi_{\epsilon , \epsilon^\prime} , u ) \, v$, implying that $\xi_{\epsilon , \epsilon^\prime}$ and $v$ are collinear. Whence, since they are both parallel, $\xi_{\epsilon , \epsilon^\prime} = c \, v$ for some $c \in \RR$. Furthermore, the R-symmetry parameter $\rho_{\epsilon , \epsilon^\prime}$ (defined above \eqref{eq:rhoLie}) vanishes identically since $\epsilon$ and $\epsilon^\prime$ are parallel. 

Thus $\cF$ is spanned by parallel spinors on a Brinkmann-wave and $[\cF,\cF]$ spans $\RR  v$, which is a central element in the conformal symmetry superalgebra. Clearly this rather trivial type of conformal Killing superalgebra associated with a Brinkmann-wave can be constructed in any dimension, solving as it does the Jacobi identity \eqref{eq:oddoddoddjacobi} in a trivial manner. In this sense, we consider it a non-generic solution relative to the analysis in section~\ref{sec:CKS}.

%%%%%%%%%%%
\subsection{Lorentzian Einstein-Sasaki manifolds}
\label{sec:LorentzianES}

From proposition 3.2 in \cite{BL:2003}, it follows that $M$ is locally isometric to a Lorentzian Einstein-Sasaki manifold (with Lorentzian Einstein-Sasaki structure $\zeta_\epsilon$)  whenever $m$ is odd and $M$ admits a twistor spinor $\epsilon$ such that

$\bullet$ $\zeta_\epsilon$ is a timelike Killing vector field. \\ [.1cm]
$\bullet$ ${\bm \zeta_\epsilon} \epsilon = \mu \epsilon$, for some non-zero constant $\mu$. \\ [.1cm]
$\bullet$ $\nabla_{\zeta_\epsilon} \epsilon = \lambda \epsilon$, for some non-zero constant $\lambda$. 

Comparing what one gets by acting with ${\bm \zeta_\epsilon}$ and $(\epsilon ,-)$ on the second condition above fixes $\mu = -i (\epsilon,\epsilon)$ to be an imaginary constant. Whence, the norm of $\zeta_\epsilon$ is $- (\epsilon,\epsilon)^2$. Acting with $(\epsilon ,-)$ on the third condition above fixes $\lambda = -\tfrac{i}{m} \varsigma_\epsilon$, in terms of the real constant $\varsigma_\epsilon$ associated with $\epsilon$ defined in \eqref{eq:DiracConstant}. 

From these properties, it follows that ${\hat \cL}_{\zeta_\epsilon} \epsilon =  ( \tfrac{m+1}{2} ) \lambda \epsilon = -i\, ( \tfrac{m+1}{2m} ) \varsigma_\epsilon \epsilon$. Whence,
\begin{equation}\label{eq:ESjacobi}
{\hat \cL}_{\zeta_\epsilon} \epsilon + i \rho_\epsilon \epsilon = 0~,
\end{equation}
provided 
\begin{equation}\label{eq:ESrho}
\rho_\epsilon = \left( \frac{m+1}{2m} \right) \varsigma_\epsilon~.
\end{equation}
For $m=3,5$, the coefficient in \eqref{eq:ESrho} matches precisely the one derived from Table~\ref{tab:tableCKS} when $N=1$. Furthermore, \eqref{eq:ESrho} prescribes a conformal symmetry superalgebra for Lorentzian Einstein-Sasaki manifolds with $m=3+4k$ and $N=1$, for any positive integer $k$.

%%%%%%%%%%%
\subsection{Fefferman spaces}
\label{sec:Fefferman}

Subsequent to their original construction within the context of CR geometry, Fefferman spaces have since been found to admit several more convenient characterisations \cite{Sparling:1985,Graham:1987,Baum:1999,BL:2003}. From proposition 3.3 in \cite{BL:2003}, it follows that $M$ is locally isometric to a Fefferman space whenever $m$ is even and $M$ admits a twistor spinor $\epsilon$ such that

$\bullet$ $\zeta_\epsilon$ is a regular null Killing vector field. \\ [.1cm]
$\bullet$ ${\bm \zeta_\epsilon} \epsilon =0$. \\ [.1cm]
$\bullet$ $\nabla_{\zeta_\epsilon} \epsilon = \lambda \epsilon$, for some non-zero constant $\lambda$. 

At least locally, there must exist a vector field $\theta$ on $M$ with $g(\theta,\zeta_\epsilon) \neq 0$. Acting with $({\bm \theta} \epsilon ,-)$ on the third condition above fixes $\lambda = -\tfrac{2i}{m} \varsigma_\epsilon$, using ${\bm \zeta_\epsilon} \epsilon =0$. 

From this it follows that ${\hat \cL}_{\zeta_\epsilon} \epsilon = ( \tfrac{m+2}{4} ) \lambda \epsilon =  -i\, ( \tfrac{m+2}{2m} ) \varsigma_\epsilon \epsilon$. Whence,
\begin{equation}\label{eq:Feffermanjacobi}
{\hat \cL}_{\zeta_\epsilon} \epsilon + i\rho_\epsilon \epsilon = 0~,
\end{equation}
provided 
\begin{equation}\label{eq:Feffermanrho}
\rho_\epsilon = \left( \frac{m+2}{2m} \right) \varsigma_\epsilon~.
\end{equation}
For $m=4,6$, the coefficient in \eqref{eq:Feffermanrho} matches precisely the one derived from Table~\ref{tab:tableCKS} when $N=1$. Furthermore, \eqref{eq:Feffermanrho} prescribes a conformal symmetry superalgebra for Fefferman spaces with $m>6$ and $N=1$. Any such conformal symmetry superalgebra is isomorphic to the conformal Killing superalgebra of a Fefferman space on which there is only one linearly independent chiral projection of any given twistor spinor (e.g. as is the case for $m=0$ mod $4$).

%%%%%%%%%%%
\subsection{Direct products}
\label{sec:DirectProducts}

Let $M = L \times R$, where $L$ is a Lorentzian Einstein-Sasaki manifold and $R$ is an irreducible non-flat Riemannian manifold which admits a Killing spinor (with imaginary Killing constant). The dimension $l$ of $L$ is odd and let $r$ denote the dimension of $R$. Any such $R$ is necessarily Einstein with non-negative scalar curvature. Recall that any Einstein three-manifold has constant curvature and is locally conformally flat. 

Let $g_R$ denote the Riemannian metric on $R$. The {\emph{metric cone}} of $R$ is the manifold $C (R) = \RR^+ \times R$ with metric $dt^2 + t^2 g_R$, where $t \in \RR^+$. The characterisation due to B\"{a}r \cite{Bar:1993} establishes that $R$ admits a Killing spinor (with imaginary Killing constant) if and only if $C(R)$ admits a parallel spinor. If $R$ is complete, it follows from \cite{Gallot:1979,Wang:1989} that ${\mathrm{Hol}}_\nabla (C(R))$ must be  either ${\mathrm{SU}}(\tfrac{r+1}{2})$, ${\mathrm{Sp}}(\tfrac{r+1}{4})$, $G_2$ (if $r=6$), $\Spin(7)$ (if $r=7$) or trivial. In these cases, $R$ is respectively Einstein-Sasakian, {\emph{3-Sasakian}}, {\emph{nearly K\"{a}hler}}, {\emph{weak $G_2$}} or a sphere. 

Thus, if $R$ is not locally isometric to a sphere then $m \geq 8$. For $m=8$, $L$ must have dimension three and $R$ must be an Einstein-Sasaki five-manifold. For $m=9$, $L$ must have dimension three and $R$ must be nearly K\"{a}hler. For $m=10$, if $L$ has dimension five then $R$ must be an Einstein-Sasaki five-manifold. Alternatively, if $L$ has dimension three then $R$ must be either Einstein-Sasakian, $3$-Sasakian or weak $G_2$.  

Let us first consider the case where $R$ has odd dimension. Since $m$ is even, with both $l$ and $r$ odd, the Clifford algebra $\Cl (m-1,1)$ can be decomposed in terms of the $\Cl (2,0) \otimes \Cl (l-1,1) \otimes \Cl (r,0)$ subalgebra in the tangent space of $M = L \times R$. This allows one to express 
\begin{equation}\label{eq:OddOddCliffordDec}
{\bm X} = \left( \begin{smallmatrix} 0&-i\\ i&0 \end{smallmatrix} \right) \otimes {\bm X_L} \otimes {\bf 1}_R + \left( \begin{smallmatrix} 0&1\\ 1&0 \end{smallmatrix} \right) \otimes {\bf 1}_L \otimes {\bm X_R}~,
\end{equation}
for all $X = ( X_L , X_R ) \in \fX (M)$. 

The $\pm$ eigenspaces of     
\begin{equation}\label{eq:OddOddCliffordChirality}
{\bm \Omega} = \left( \begin{smallmatrix} 1&0\\ 0&-1 \end{smallmatrix} \right) \otimes {\bf 1}_L \otimes {\bf 1}_R~,
\end{equation}
contain chiral projections of the form 
\begin{equation}\label{eq:ChiralSpinors}
\epsilon_+ = \left( \begin{smallmatrix} 1\\ 0 \end{smallmatrix} \right) \otimes \epsilon_L \otimes \epsilon_R \;\; , \quad\quad \epsilon_- = \left( \begin{smallmatrix} 0\\ 1 \end{smallmatrix} \right) \otimes \epsilon_L \otimes \epsilon_R~,
\end{equation}
for all $\epsilon \in \fS (M)$, where $\epsilon_L \in \fS (L)$ and $\epsilon_R \in \fS (R)$ are both Killing spinors. If $\nabla_{X_L} \epsilon_L = \mu \, {\bm X_L} \epsilon_L$, for all $X_L \in \fX(L)$ and some $\mu \in \RR$, then it is straightforward to check that $\epsilon_\pm \in \fS_\pm^c (M)$ only if $\nabla_{X_R} \epsilon_R = \pm i \mu \, {\bm X_R} \epsilon_R$, for all $X_R \in \fX(R)$. Whence, $\epsilon_\pm$ cannot both be twistor spinors on $M$ unless $\mu =0$. Henceforth, we shall take $\epsilon_+ \in \fS_+^c (M)$ with $\mu \neq 0$.  

The properties above imply

$\bullet$ $\zeta_{\epsilon_+}$ is a null Killing vector field. \\ [.1cm]
$\bullet$ ${\bm \zeta_{\epsilon_+}} \epsilon_+ =0$. \\ [.1cm]
$\bullet$ $\nabla_{\zeta_{\epsilon_+}} \epsilon_+ = 2i\mu\, ( \epsilon_L,\epsilon_L) (\epsilon_R,\epsilon_R) \epsilon_+$. 

From this it follows that ${\hat \cL}_{\zeta_{\epsilon_+}} \epsilon_+ = ( \tfrac{m+2}{4} ) \nabla_{\zeta_{\epsilon_+}} \epsilon_+ =  -i\, ( \tfrac{m+2}{2m} ) \varsigma_{\epsilon_+} \epsilon_+$. Whence,
\begin{equation}\label{eq:Productjacobi}
{\hat \cL}_{\zeta_{\epsilon_+}} \epsilon_+ + i\rho_{\epsilon_+} \epsilon_+ = 0~,
\end{equation}
provided 
\begin{equation}\label{eq:Productrho}
\rho_{\epsilon_+} = \left( \frac{m+2}{2m} \right) \varsigma_{\epsilon_+}~.
\end{equation}
For $m=4,6$, the coefficient in \eqref{eq:Productrho} again matches precisely the one derived from Table~\ref{tab:tableCKS} when $N=1$. Furthermore, \eqref{eq:Productrho} prescribes a conformal symmetry superalgebra for higher-dimensional direct product spaces $M = L \times R$ with $r$ odd and $N=1$.

If $R$ has even dimension then it must be either nearly K\"{a}hler or an even-dimensional sphere. Since $m$ is odd, with $l$ odd and $r$ even, the Clifford algebra $\Cl (m-1,1)$ can be decomposed in terms of the $\Cl (l-1,1) \otimes \Cl (r,0)$ subalgebra in the tangent space of $M = L \times R$. This allows one to express 
\begin{equation}\label{eq:OddEvenCliffordDec}
{\bm X} = {\bm X_L} \otimes {\bm \Omega_R} + {\bf 1}_L \otimes {\bm X_R}~,
\end{equation}
for all $X = ( X_L , X_R ) \in \fX (M)$. Moreover,
\begin{equation}\label{eq:OddEvenSpinors}
\epsilon = \epsilon_L \otimes \epsilon_R~,
\end{equation}
for all $\epsilon \in \fS (M)$, where $\epsilon_L \in \fS (L)$ and $\epsilon_R \in \fS (R)$ are both Killing spinors. If $\nabla_{X_L} \epsilon_L = \mu \, {\bm X_L} \epsilon_L$, for all $X_L \in \fX(L)$ and some $\mu \in \RR$, then it is straightforward to check that $\epsilon \in \fS^c (M)$ only if $\nabla_{X_R} \epsilon_R = \mu \, {\bm X_R} {\bm \Omega_R} \epsilon_R$, for all $X_R \in \fX(R)$. This condition on $\epsilon_R$ is equivalent to $\epsilon_R^+ \pm i \epsilon_R^-$ being Killing spinors, with imaginary Killing constants $\pm i\mu$, where $\epsilon_R^\pm = \half ( {\bf 1}_R \pm {\bm \Omega_R} ) \epsilon_R$.

The properties above imply

$\bullet$ $\zeta_{\epsilon}$ is a null Killing vector field. \\ [.1cm]
$\bullet$ ${\bm \zeta_{\epsilon}} \epsilon =0$. \\ [.1cm]
$\bullet$ $\nabla_{\zeta_{\epsilon}} \epsilon = 2i\mu\, ( \epsilon_L,\epsilon_L) (\epsilon_R,\epsilon_R) \epsilon$. 

From this it follows that ${\hat \cL}_{\zeta_{\epsilon}} \epsilon = ( \tfrac{m+2}{4} ) \nabla_{\zeta_{\epsilon}} \epsilon =  -i\, ( \tfrac{m+2}{2m} ) \varsigma_{\epsilon} \epsilon$. Whence,
\begin{equation}\label{eq:Productjacobi2}
{\hat \cL}_{\zeta_{\epsilon}} \epsilon + i\rho_{\epsilon} \epsilon = 0~,
\end{equation}
provided 
\begin{equation}\label{eq:Productrho2}
\rho_{\epsilon} = \left( \frac{m+2}{2m} \right) \varsigma_{\epsilon}~.
\end{equation}
In this case, \eqref{eq:Productrho2} prescribes a conformal symmetry superalgebra for higher-dimensional direct product spaces $M = L \times R$ with $r$ even and $N=1$.

%%%%%%%%%%%%%%%%%%%%%%

\section*{Acknowledgments}

It is a pleasure to thank Jos\'{e} Figueroa-O'Farrill for some useful comments. The financial support provided by ERC Starting Grant QC \& C 259562 is gratefully acknowledged.

\bibliographystyle{utphys}
\bibliography{CKS}

\end{document}